\documentclass[acus]{JAC2003}

\usepackage{graphicx}
\usepackage{subcaption} %  for subfigures environments
\usepackage{booktabs}
\usepackage{pdflscape}
\usepackage{textcomp}
\begin{document}
\title{LLRF System Analysis for the Fermilab PIP-II Superconducting LINAC
\thanks{The authors of this work grant the arXiv.org and LLRF Workshop's International Organizing Committee a non-exclusive and irrevocable license to distribute the article, and certify that they have the right to grant this license.}\thanks{ Work supported by FermiForward Discovery Group, LLC under Contract No. 89243024CSC000002 with U.S. DOE.}}

\author{P. Varghese\thanks{ varghese@fnal.gov},S. Raman, M. Guran, L. Reyes \\
Fermi National Accelerator Laboratory (FNAL), Batavia, IL 60510, USA \\
L. Doolittle (Retd.), Q. Du,  S. Murthy \\
Lawrence Berkeley National  Laboratory (LBNL), Berkeley, CA 94720, USA}
\maketitle

\begin{abstract}
   PIP-II is a superconducting linac that is in the initial acceleration chain for the Fermilab accelerator complex. The RF system consists of a warm front-end with an RFQ and buncher cavities along with 25 superconducting cryo-modules comprised of cavities with five different acceleration \(\beta\). The LLRF system for the linac has to provide field and resonance control for a total of 125 RF cavities. Various components of the LLRF system have been tested with and without beam at the PIP-II test stands. The LLRF system design is derived from the LCLS-II project with its self-excited loop architecture used in the majority of the cryo-modules. The PIP-II beam loading at 2 mA is much higher than the LCLS-II linac. The control system architecture is analyzed and evaluated for the operational limits of feedback gains and their ability to meet the project regulation requirements for cavity field amplitude and phase regulation.  
\end{abstract}

\section{INTRODUCTION}
\par
The PIP-II project at Fermilab is a new superconducting linac feeding the existing Booster, Recycler and Main Injector accelerator rings enabling them to provide a 1.2 MW proton beam
over the energy range of 60-120 GeV to drive neutrino research at the Deep Underground Neutrino Experiment (DUNE) and for the Mu2e project.
The LLRF system for PIP-II is based on the LCLS-II LLRF system design with the same collaboration labs participating in the effort. The system design was lead by the team from LBNL who also contributed the bulk of the firmware/software codebase. The design uses the SEL architecture pioneered by J. Delayen and was first implemented in a digital FPGA platform at JLAB[1,2]. The current implementation for LCLS-II and PIP-II uses an architecture where the FPGA is used primarily to implement the signal processing and control loops while the bulk of the numerical calculations during calibration and other computations is performed in an external server using numerous Python scripts. Thus the details of the overall control system are scattered across dozens of python scripts and several dozens of HDL code modules that are not fully documented. The team from Fermilab has worked closely with the LBNL team to document various components of the codebase with a focus on the main signal chain, the key calibrations and the SEL architecture implementation. Following the documentation effort, it became possible to take a closer look at how the system will perform for the PIP-II linac. While there are many similarities with the LCLS-II linac, one of the key differences is the level of beam loading which is greater by a factor of 20. As part of the feedback system performance analysis, some of the implementation details are described to provide a basis for the conclusions reached. The modifications needed to the firmware/software to improve performance are also explored.   
\section{Self Excited Loop Fundamentals}
\par
The basic configuration of the  self excited loop architecture is shown in Fig. 1. The first cordic in the signal chain converts the I,Q inputs to magnitude and phase followed by their respective feedback blocks. However, the outputs of the feedback blocks are not converted back directly to I,Q components.   The key feature of this architecture is that the amplitude and phase feedback
loop outputs are treated as I and Q inputs that are rotated by the sum of the measured
phase and the phase offset that makes the self excited loop oscillation possible by forming a positive feedback configuration in the cavity drive loop. This is essentially a 
digital equivalent of the original analog self-excited loop configuration[1]. When the cavity frequency is on resonance, the phase feedback  loop set point is near zero 
representing an acceleration phase angle near the crest of the cavity field. The I and Q inputs to the second cordic drive the real and imaginary components of the cavity
forward power. The configurable saturation components and the limits they are set to for different operational modes are an essential feature of the SEL architecture. In order to
compute these limits for various cavity field settings, some power and amplifier calibrations are required, which will be described next.
\begin{figure}[!h]
\centering
\includegraphics[width=3.0in]{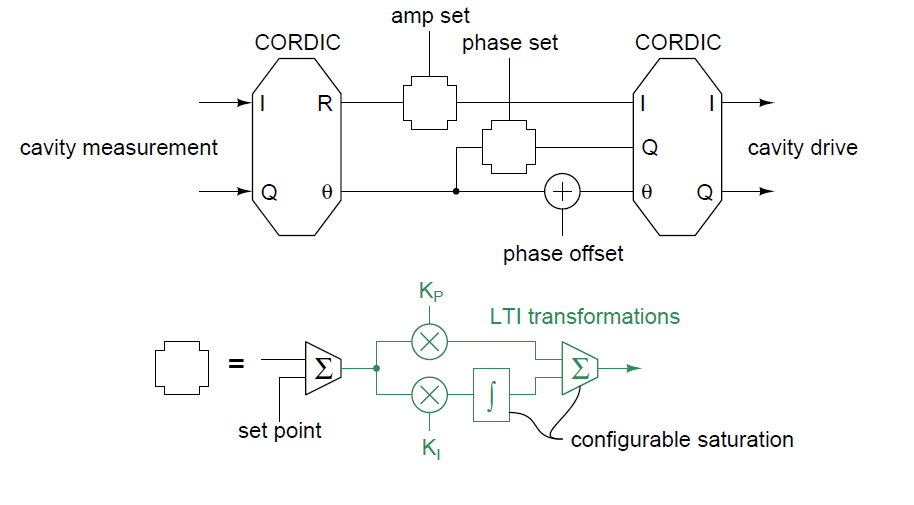}
\caption{LLRF System Architecture - SEL Control}
\label {fig1}
\end{figure}

\subsection {Signal Calibration}
\par
A 10 dBm RF signal is input to the downconverter channels for forward, reflected and cavity probe input and the corresponding digitizer channel
ADC counts are recorded as representing the maximum input power values for those signals. The separately measured cable attenuations, directional coupler attenuation and any
additional attenuators used in the signal path are summed and the resultant is used to compute the power levels at the cavity.
\subsection {Gradient Calibration}
There are two methods used for gradient calibration - one based on transmitted power and a more accurate one based on reverse power measurement
during the cavity field decay following a pulse power input. The former meaurement is similar to the signal calibration method. With a 10 dBm input at the 
downconverter input for the cavity probe, the ADC counts are recorded. Using the cable attenuation measurement value, the probe power for full scale
ADC reading is calculated.
\par
The reverse power gradient calibration method calculates the cavity stored energy U, by computing the area under the
reverse power decay curve when the cavity drive pulse is turned off in a cavity pulse calibration. The cavity voltage can be calculated
using the relationship
\begin{equation}   
P_{diss} = {{V_0}^2 \over R} \quad  = {{\omega_0 \times U} \over {Q}}
\end{equation}
Solving for the cavity voltage
\begin{equation}   
V_0 = {\left[ U \times {r \over Q} \times 2 \pi f_0\right] }^{1 / 2}
\end{equation}

\begin{figure*}[!t]
\centering
 \includegraphics[width=7in]{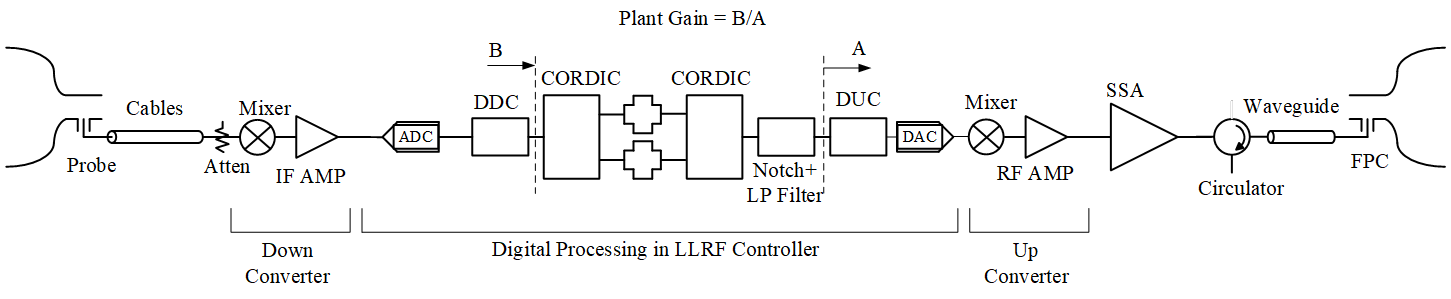}
\caption{LLRF Feedback Loop Components}
\label {fig3}
\end{figure*}

\subsection {SSA Calibration}
\par
SSA calibration is an essential measurement that is required for the SEL modes of operation. It establishes a mathematical model of the cavity forward power 
to the DAC drive level. A signal calibration described in the previous section must be completed prior to the SSA calibration. This relationship is usually approximated by a linear fit passing through the origin for the amplifier power characteristic. This 
linear fit results in a SSA slope that models the amplifier output for any DAC setting. An example of a plot of this calibration is shown in Fig. 3.  
\begin{figure}[!b]
\centering
 \includegraphics[width=3.0in]{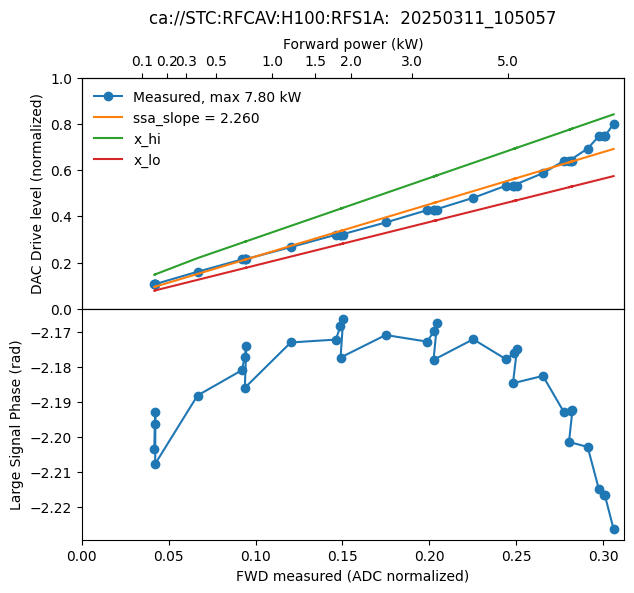}
\caption{SSA Calibration}
\label {fig3}
\end{figure}

\section{SEL Modes}
\par
There are four modes of operation associated with the SEL architecture - SEL Raw, SEL, SELA and SELAP. The first two are effectively open loop 
from the feedback point of view although there is no switch that actually opens the loop. The mode change is made by adjusting the drive limits on the
output side of the amplitude and phase feedback loops. Setting both the upper and lower limit to be the same fixed number in the saturation blocks, ignores the output
from the feedback loops, thus making the configuration effectively open loop. The SELA mode has the amplitude feedback loop closed with the phase loop open.
The SELAP mode has both amplitude and phase loops in feedback mode with the saturation limits set as indicated in Table 1. 
\begin{table}[!t] 
\caption{Saturation Limits for SEL Modes}
\label{tab:sample}
%\vspace{8pt}
\centering

\begin{center}
    \begin{tabular}{ | c | c | c | c | c | }
    \hline
%%\bfseries Mode & \bfseries x_low ({\%} & \bfseries x_hi) & \bfseries y_low & \bfseries y_high \\ \hline
Mode & \(X_{lo}\)  & \( X_{hi}\) & \(Y_{lo}\) &  \(Y_{hi}\) \\ \hline
SEL Raw & - & - & - & -  \\ \hline 
SEL & \( X_{tgt}\) & \( X_{tgt}\) & 0 & 0  \\ \hline
SELA &  \( X_{tgt}\times0.85\) & \( X_{tgt}\times1.15\) & 0 & 0  \\  \hline
SELAP & \( X_{tgt}\times0.85\) & \( X_{tgt}\times1.15\) & \(- Y_{p}\) & \(Y_{p}\)   \\ \hline
    \end{tabular}
\end{center}
\end{table}
The X and Y parameters refer to the I and Q inputs to the second cordic shown in Fig. 1.
 The computation of \(X_{tgt}\) and \(Y_p\) involves the signal calibration and SSA calibration results as will be shown here.
\par
Let \(E_{Des}\) be the requested gradient. Then
\begin{equation}   
V_{Des} = E_{Des} \times l \quad MV
\end{equation}
where l is the cavity length. The stored energy is computed from Eq. (2).
\begin{equation}   
{\sqrt{U} } = {{V_{Des}} \over {\left[ {r \over Q} \times {2 \pi f_0} \right]^{1 / 2} }} \quad {\sqrt{J} }
\end{equation}
 The required foward power is given by
\begin{equation}   
\sqrt{P}  = \sqrt {U} \times \sqrt{{\pi f_0} \over  {2 Q_L}} \quad \sqrt{W} 
\end{equation}
The corresponding normalized forward power ADC setting is
\begin{equation}   
ADC_{Fwd}  = {\sqrt {P} \over C_{FSfwd}}
\end{equation}
where \(C_{FSfwd}\) is the forward power calibration constant.
This represents the X co-ordinate of the SSA calibration curve fit. The corresponding normalized DAC drive is \(X_{tgt}\) in Table 1.
The +/- 15\%  lines about the linear fit on the SSA calibration represent the \(X_{hi}\) and  \(X_{lo}\) limits.
\begin{equation}   
X_{tgt}  = SSASlope \times ADC_{Fwd}
\end{equation}
\par
The saturation limits of the phase loop that is represented by \(Y_{hi}\) and  \(Y_{lo}\) follow a different computational path. There are two
 user settable parameters that affect the limits of the phase loop. The first is a DAC normalized drive limit \(D_{Max}\) that is settable
from 0 - 1. The second parameter \(D_{Imag}\) represents the reactive component of the drive power. The real component 
of the drive power determines the stored energy and therefore the cavity field. The imaginary component compensates for cavity detuning due to 
microphonics or other disturbances. Thus the phase limits are computed as

\begin{equation}   
D_{Real}  = \sqrt {1 - {D_{Imag}}^2}
\end{equation}
\begin{equation}   
Y_p  = D_{Max} \times D_{Imag}
\end{equation}
\begin{equation}   
X_{Max}  = D_{Max} \times D_{Real}
\end{equation}
The real part of the drive is limited to \(X_{Max}\) in all modes.
\begin{figure*}[!bt]
\centering
  \includegraphics[width=7in]{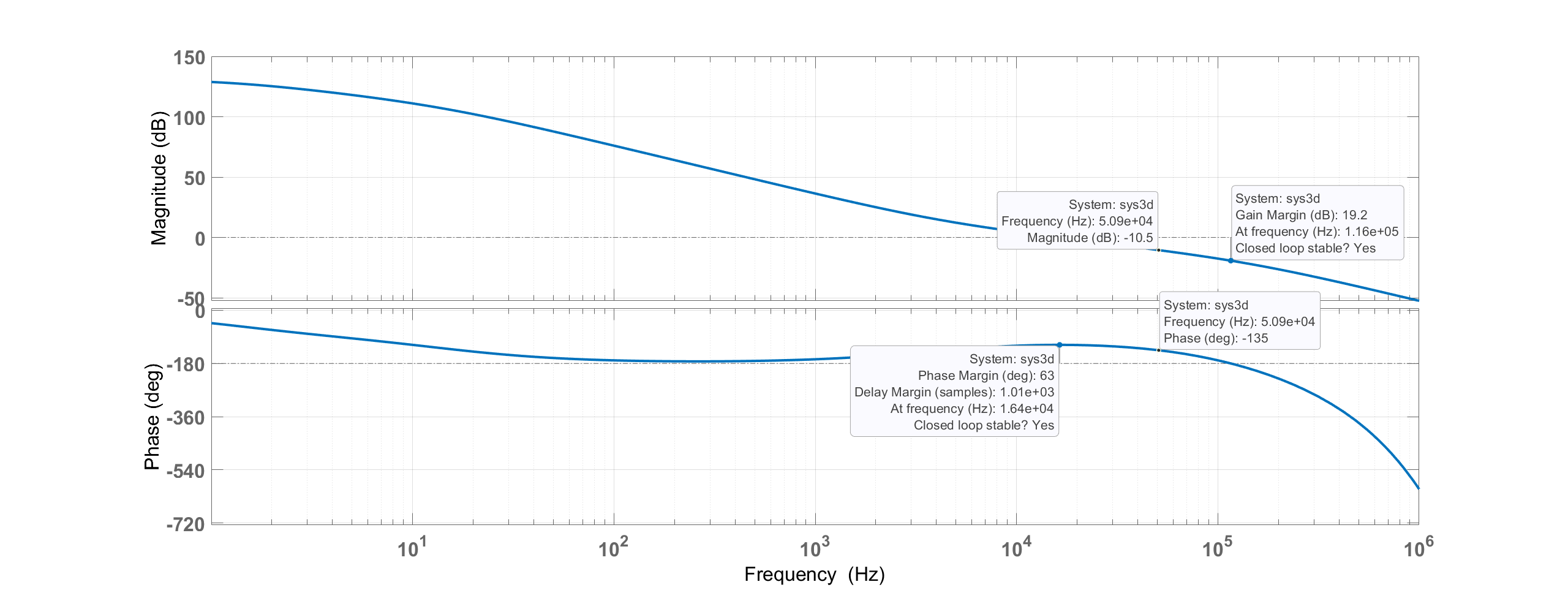}
\caption{LCLS-II Cavity Bode Analysis}
\label {fig7}
\end{figure*}

\section{Stability Analysis}
\par
The current criterion for selecting the optimal gain is determined by the limits of the excursions of phase and amplitude to within
a 'box' defining the limits of reactive power and real power components of the forward drive signal. Output limiting is a necessity in the SEL configuration due to the positive feedback
nature of the architecture. However, when the amplitude and phase loops are closed in a negative feedback configuration as expalined in the previous section, the limiting function is
being carried out by the stability of the feedback loop. Under these conditions, the stablity of the feedback loop can be analyzed using the bode frequency response analysis method.
\par
The bode plot of the LCLS-II system with some nominal cavity and field parmeters, is shown in Fig. 4. The gain margin with a 45 degree phase margin is 10.5 dB. The assumed system gain in the plot is 1000. Thus the maximum gain with a 45 degree phase margin is given by

\begin{equation}   
Sys_{Pgainmax} =1000 \times 10^ {10.5/20 }  \quad  = 3350 
\end{equation}
where a loop delay of 1.2 \(\mu sec\) is assumed.
If the loop delay is increased to 2.2 \(\mu sec\) which is probably closer to the overall loop delay, max gain reduces to 2075. This is close to the maximum value the software limits it to,  of 2048.
The corresponding integral gain limits in the software are 47e6 rads/sec.
\par
\begin{figure}[!t]
\centering
 \includegraphics[width=3.0in]{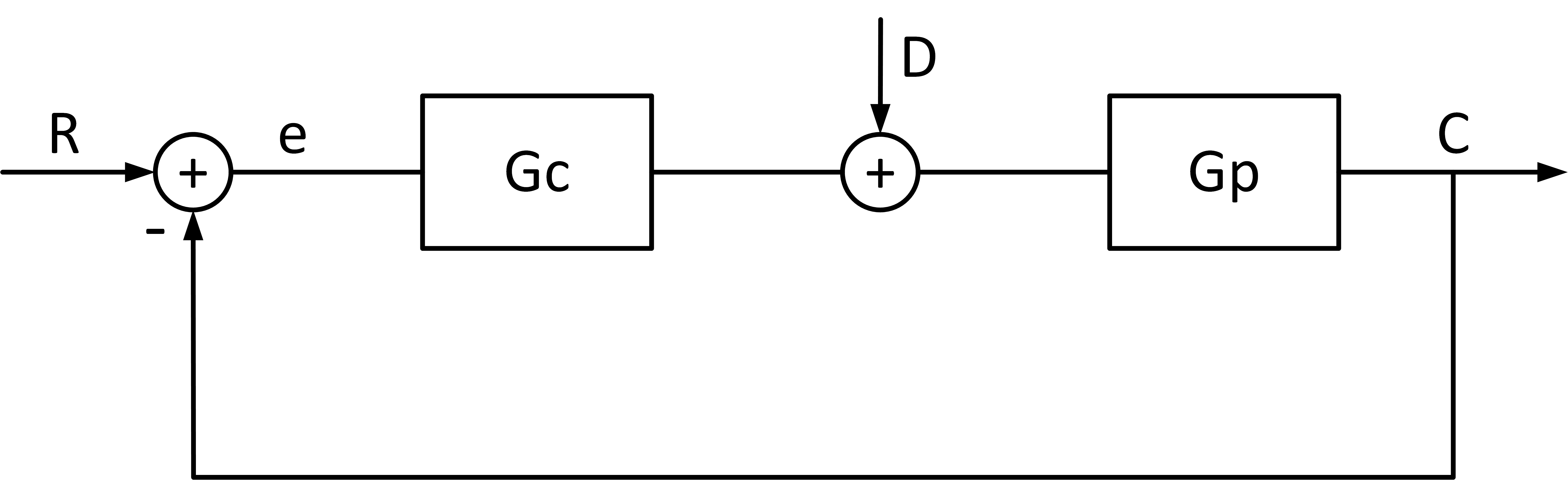}
\caption{Feedback Regulation}
\label {fig6}
\end{figure}

Consider the basic feedback loop configuration shown in Fig. 5. Feedback performance is specified in terms of percentage rms value of the error e. In the presence of an external disturbance D, the contribution to the feedback error from the disturbance is given by
\begin{equation}   
{\left|e\right| } = {{\left|D\right| } \over {1 + G_c}}
\end{equation}
where \(G_p\) is assumed to be 1 for simplicity. In order to meet a regulation specification, the system gain has to be sufficiently high for a given disturbance level. Conversely, if the system is operated at lower gains, the disturbance levels need to be low enough to satisfy system performance specifications.
 
\section{Performance Limitations}
\par
LLRF system performance is usually specified in terms of cavity field amplitude and phase regulation. For the LCLS-II project these specifications are 0.01 \(\% \) rms and 0.01 degrees rms, for amplitude and phase respectively. The corresponding numbers for the PIP-II project are 0.065 \(\%\) rms and 0.065 degrees rms. In this section we will analyze the capability of the control system to meet these specifications and if feedback gains are limited by the control architecture used, to determine what the achievable performance limits are. 
\subsection {Amplitude Loop}
\par
The amplitude and phase loops are independent in the SEL architecture. Therefore, the performance of each loop will be examined separately. Consider the SSA calibration and cavity drive box limits shown in Fig. 3. The cavity field amplitude A is proportional to the square root of the real component of the drive power. Let \(\delta A\) be the maximum amplitude disturbance before the box limits
are reached. Then
\begin{equation}   
A = K_C \sqrt {P_N}  ,  \quad  A + \delta A = K_C \sqrt{P_N + \delta P} 
\end{equation}
Since the upper saturation limit for real power is 115 \(\%\) of the nominal power \(P_N\)
\begin{equation}   
{P_N + \delta P} = 1.15 \times P_N
\end{equation}
 Taking the ratio of the two expressions in Eq. (40), we can write
\begin{equation}   
1 + {{\delta A} \over {A}} = \sqrt {1.15}  ,  \quad  { \delta A  \over A}= 0.0724
\end{equation}
When the above limits are applied, the minimum gain needed to meet the specifications and the corresponding maximum disturbance limits are
 \begin{equation}   
G_{Min} = {{0.0724} \over {.0006}} = 120 , \quad  {D \over A}  \le  0.0724
\end{equation}
Stability analysis based on PIP-II SSR1 cavity controller and loop delays(\(\approx 3 \mu \)S) gives a maximum gain of 880. With this maximum gain, the corresponding
maximum disturbance level is given by
\begin{equation}   
G_{Max} = 880  , \quad    {D_{Max} \over A}  \le  {0.528} = 880 \times 0.0006
\end{equation}
provided the box limits are increased.
\subsection{Phase Loop}
\par
The phase loop saturation limits are set differently as outlined in Eq. (9). The user enters a maximum drive limit and a reactive power fraction. This number is typically set around 0.6 which implies a real power fraction of 0.8. The reactive power limits are shown in Fig. 6 as the vertical axis limits of the box. According to the SEL principle, the reactive power changes to keep the cavity in resonance to compensate for the microphonic disturbances that cause the cavity to detune. The reactive fraction of 0.6 (\(sin \theta\)) results in a phase rotation of about +/- 37 degrees.  To maintain a phase regulation of better than 0.06 degrees, the phase gain needs to be greater than 583.3 if the disturbances are driving the reactive power to the limits. This is computed as
\begin{equation}   
G_{Min} ={{37} \over {.06}} \approx  583 , \quad  D_{\Phi}  \le  37 \quad {degrees}
\end{equation}
Using the maximum gain for stability of the SSR1 cavity again we get an upper limit for the phase disturbance
\begin{equation}   
G_{Max} = 880  , \quad    D_{\Phi}   \le  {52.8} = 880 \times 0.06 \quad {degrees}
\end{equation}
which corresponds to a reactive fraction mximum of
\begin{equation}   
D_{ImagMax} = sin(52.8) = 0.8
\end{equation}

 Regardless of what the reactive fraction value is, it is the actual disturbances present that determine the minimum gain in the phase loop to meet phase regulation specifications. 
\begin{table*}[!t]
\caption{Beam Loading in PIP-II Cavities}
\label{tab:sample}
%\vspace{8pt}
\centering

\begin{center}
    \begin{tabular}{ | c | c | c | c | c | c | }
    \hline
%%\bfseries Mode & \bfseries x_low ({\%} & \bfseries x_hi) & \bfseries y_low & \bfseries y_high \\ \hline
Cavity & Volts(MV)  & Fwd NB(kW) &  Fwd WB(kW) &  \(\sqrt{Pwr Ratio}\) & Amp Max(kW)\\ \hline
HWR-6 & 2.008 & 2.64 & 4.45 & 1.3 & 7  \\ \hline
SSR1-8 & 2.050 & 1.98 & 4.13 & 1.44 & 7  \\ \hline
SSR2-5-4 & 4.993 & 6.4 & 11.92 & 1.36 & 20  \\ \hline
LB650-5-3 & 11.88 & 15.93 & 28.97 & 1.35 & 40  \\ \hline
HB650-4-2 & 19.95 & 24.28 & 40.71 & 1.29 & 70  \\ \hline  
    \end{tabular}
\end{center}
\end{table*}
\begin{figure}[!ht]
\centering
  \includegraphics[width=2.5in]{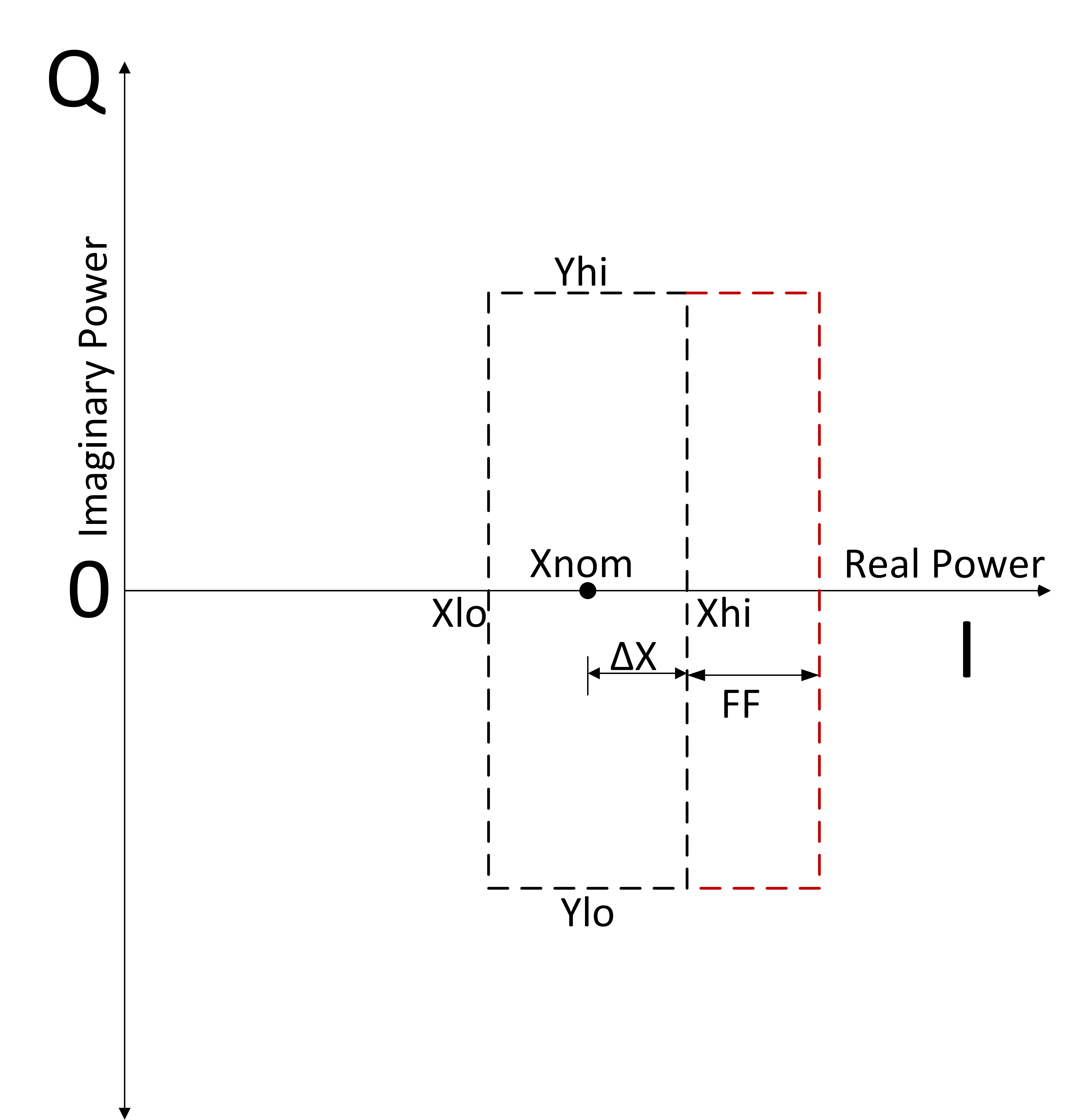}
\caption{Saturation Limits for Real and Imaginary components of drive power}
\label {fig9}
\end{figure}

\section{Beam Loading}
Beam loading is a large disturbance that could exceed these limits placed on the RF drive. The PIP-II project beam loading (2mA) is much larger compared to LCLS-II and will require these limit calculations to be expanded to give the feedback more headroom to maintain field regulation. Table. 2 shows the beam loading effects on each PIP-II cavity type with the squareroot of the ratio of the forward power with and without beam in  column 5. This is clearly greater than the 1.15 factor currently used with LCLS-II and will need to be increased unless feedforward beam loading compenstaion is applied. If the correct level of beam loading compensation is applied in a gated manner during the beam interval which has an \(\approx 1 \%\) duty cycle, then the current power limits can probably be left at \(\pm 15\%\). The feedforward addition must be placed after the saturation bocks for amplitude and phase.
\par
The two cordics inherently required in the SEL implementation  increase the loop delay for the feedback system. However the current implementation makes this problem worse by sacrificing compute cycles for resource optimization. With improvements to the main signal processing chain in the firmware, it should be possible to reduce this delay by a factor of 2 giving a corresponding increase in feedback gains by a factor of two. The higher gain will allow a more robust operation of the two feedback loops with improved disturbance rejection. 

\section {Conclusion}
\par The FPGA implementation details of the SEL LLRF architecture used in the LCLS-II and PIP-II projects is studied to extract the limits of performance of the control system for PIP-II. The system architecture places saturation limits on the output drive. This imposes reduced gain settings on the control system which degrades its ability to maintain regulation performance in the presence of disturbances such as beam loading. The SELAP mode of operation provides for minimizing the effects of the cavity detuning by increasing the quadrature power component. When the actuator saturation limits are crossed, feedback is no longer able to  maintain the field amplitude and phase at their setpoints. This results in the control system dropping out of the SELAP mode resulting in holding off the beam, till the drive  limits are no longer being crossed. The impact of the output limiting on the controllers response to the higher beam loading conditions for the PIP-II linac were evaluated. Improvements to the controller firmware such as adding feedforward and reducing latency have already commenced and the control system is expected to perform significantly better by the time the PIP-II Linac is ready for beam by 2030.


\begin{thebibliography}{5}   % Use for  1-9  references
%\begin{thebibliography}{99} % Use for 10-99 references
\bibitem{}J. R. Delayen, “Phase and Amplitude Stabilization of Superconducting Resonators,” Ph.D. Thesis, Caltech (1978).
\bibitem{}J. Delayen, T. Allison, C. Hovater, J. Musson, and T. Plawski,"Development of a Digital Self-Excited Loop for Field
Control in High-Q Superconducting Cavities”, in Proc. PAC’07, Albuquerque, NM, USA, Jun. 2007.
\bibitem{}L. Doolittle et. al.,"High Precision RF Control for SRF Cavities in LCLS-II", SLAC-PUB-17108, 2017
\bibitem{}L. Doolittle, S. Murthy "Data flow for LCLS-II LLRF cavity amplitude setting", LBL  LLRF Note, August 2025
\bibitem{}L. Doolittle, S. Murthy, Q. Du, PIP-II LLRF Collaboration Documentation Discussions, 2024-2025 
\bibitem{}P.Varghese et. al., "SEL Control Architecture Implementation in an FPGA," FNAL Technical Note, 2025
\bibitem{}P. Varghese, S. Raman, M. Guran, L. Reyes, "PIP-II LLRF Firmware/Software Documentation", FNAL Technical Note, 2025 
\end{thebibliography}
\end{document}